\documentclass[10pt,letterpaper,twocolumn]{article} %% two column, final layout

\usepackage{ol2}
\usepackage[draft]{hyperref}
\usepackage{amsmath}

\usepackage{bm}
\usepackage{graphics} 
\usepackage{graphicx}  
\usepackage{latexsym}                % to get LASY symbols
\usepackage{amsmath}     
\usepackage{amsfonts}  
\usepackage{amssymb}
\usepackage{amsthm}
\usepackage{dcolumn}
\usepackage{color}
\usepackage{mathrsfs}
\usepackage[left]{lineno}
\usepackage{dcolumn}% Align table columns on decimal point

\renewcommand{\phi}{\varphi}
\renewcommand{\>}{\right \rangle}

\newcommand{\ket}[1]{\left |#1\>}

\newcommand{\be}{\begin{equation}}
\newcommand{\ee}{\end{equation}}
\newcommand{\bea}{\begin{eqnarray}}
\newcommand{\eea}{\end{eqnarray}}

\begin{document}

\twocolumn[ %% activate for two-column option

\title{A Faithful Communication Hamiltonian in Photonic Lattices}

%% For REVTeX it is possible to automate superscript and e-mail callouts with the superscriptaddress option; see REVTeX4 documentation.

\author{Matthieu Bellec,$^{1}$ Georgios M. Nikolopoulos,$^{1,*}$ and Stelios Tzortzakis$^{1,2}$}

\address{
$^1$Institute of Electronic Structure \& Laser, FORTH, P.O.Box 1385, GR-71110 Heraklion, Greece \\
$^2$Department of Materials Science and Technology, University of Crete, P.O. Box 2208, GR-71003 Heraklion, Greece \\
$^*$Corresponding author: nikolg@iesl.forth.gr
}

\begin{abstract}
Faithful communication is a necessary precondition for large scale all-optical networking and quantum information processing. Related theoretical investigations in different areas of physics have led to various proposals in which finite discrete lattices are used  as channels for short-distance communication tasks. Here, in the framework of femtosecond-laser-written waveguide arrays, we present the first experimental realization of such a channel with judiciously engineered couplings. 
\end{abstract}


 ] %% activate for two-column option

%\noindent{\em Introduction.---} 
\noindent Photonic lattices (PLs) i.e., arrays of evanescently coupled waveguides, offer a remarkably simple and versatile tool for rigorous and transparent testing of various models associated mainly with tight-binding Hamiltonians~\cite{Christodoulides2003,Szameit2010a,Longhi2009}. Numerous phenomena  encountered in various areas of physics, such as Bloch oscillations~\cite{Trompeter2006}, Anderson localization~\cite{Schwartz2007}, Glauber-Fock states displacement~\cite{Keil2011c}, have been demonstrated and  studied experimentally in the context of linear PLs. Nonlinear PLs  offer the prospect of creating and controlling optical discrete solitons~\cite{Lederer2008a} and filaments~\cite{Bellec2011}. Besides their versatility, PLs offer scalability and  compatibility with the widespread silica technologies. Hence, they are expected to play a pivotal role in the route towards all-optical networking~\cite{Keil2011a} and large scale quantum information processing~\cite{Politi2008}. The faithful transfer of signals is a necessary requirement for further developments in these directions, and has thus attracted considerable interest. Most of the proposed solutions, however, rely on boundary-free signal propagation, which implies that lattice truncation and edge effects will unavoidably lead to a considerable distortion of the transmission~\cite{Longhi2010}. 
One way to circumvent such diffraction problems is to prevent the uncontrollable spread of the wavepacket, by using discrete solitons as information carriers; an approach that requires nonlinear PLs and large intensities~\cite{Christodoulides2003,Keil2011a}. For linear discrete PLs one may resort to  the segmentation of appropriate lattice sites~\cite{Keil2012}, or to the engineering of judicious coupling constants between adjacent sites~\cite{Longhi2010,Gordon2004,Joglekar2011a}. The latter scenario has been also studied thoroughly in the context of quantum networks~\cite{KAY2010}, and various faithful-communication (FC) Hamiltonians have been proposed. 

Taking advantage of the  versatility of PLs, we present the first proof-of-principle experiment on the FC Hamiltonian proposed in~\cite{Gordon2004,Nikolopoulos2004}, for a sufficiently large number of sites so that coupling engineering is necessary. Our observations are compared to theoretical predictions, and the influence of disorder and losses is discussed. 

%{\em Theory.---}
The Hamiltonian describing the dynamics of  a single excitation in a one-dimensional lattice within the tight-binding and nearest-neighbour (NN) approximations is of the form $(\hbar =1)$ 
\begin{subequations}
\label{PST_ham}
\bea
\hat{{\mathscr H}} = \sum_{k=1}^{N-1}C_{k,k+1} (\hat{a}^\dag_{k} \hat{a}_{k+1}+\hat{a}^\dag_{k+1} \hat{a}_{k}),
\label{PST_ham1}
\eea
where $\hat{a}^\dag_{k}$ is the creation operator for an excitation on the $k$th site. The $N$ sites are assumed to be on-resonant, and $C_{k,l}$ is the coupling between the sites with indices $k$ and $l$. 
As has been shown in~\cite{Gordon2004,Nikolopoulos2004}, when the couplings in the Hamiltonian (\ref{PST_ham1}) are chosen 
according to
\be
C_{k,k+1}=C_0 \sqrt{(N-k)k}, 
\label{Coup}
\ee
\end{subequations}
the lattice operates as a perfect quantum channel, i.e.,  ideally perfect transfer of the excitation from the $k$th to the $(N-k+1)$th site of the lattice  at time $\tau=\pi/(2C_0)$. 

\begin{figure}[h]
\includegraphics[width=8.3cm]{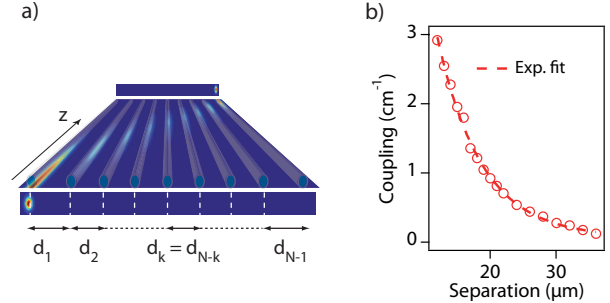}
\caption{\label{fig1} (a) Schematic view of the PL. (b) Measured dependence of the coupling constant on the waveguide separation at 633 nm and the exponential fit.}
\end{figure}

The photonic realization of Hamiltonian (\ref{PST_ham}) relies on the fact that the temporal evolution of a single excitation in the Hilbert space spanned by $\{\ket{k}\equiv\hat{a}^\dag_{k}\ket{{\bf 0}}\}$, is isomorphic to the spatial propagation of light in a linear array of evanescently coupled  single-mode waveguides, when one of the waveguides is initially excited. Within the coupled-mode theory, the light evolution in such a structure, is described by the set of discrete equations
\begin{equation}
i \dfrac{d\psi _{k}}{dz} = C_{k,k-1}\psi _{k-1} + C_{k,k+1}\psi _{k+1},
\label{eq2}
\end{equation}
where $\psi_k(z)$ is the normalized modal electric-field amplitude for the $k$th waveguide, and $C_{k,k\pm 1}$ are determined by the evanescent overlap between the transverse components of the field  modes in adjacent waveguides, whereas  the evolution of the excitation takes place in space ($z$) instead of time. In a PL of $N$ waveguides of fixed length $L$, faithful (ideally perfect) power transfer from the $k$th to the $(N-k+1)$th waveguide can be achieved if $
\{C_{k,k\pm 1}\}$ are chosen according to (\ref{Coup}), with $C_{0} = \pi /(2L)$.

In the weak-coupling regime, $C_{k,k+ 1}$ depends exponentially on the separation of the waveguides i.e., $C_{k,k+ 1}^{\rm exp} = \alpha \exp \left(-\beta d_{k,k+ 1}\right)$ \cite{Szameit2010a}, where $\alpha,\beta$ are open parameters to be determined by fitting to related experimental data for a particular setup. For given $N$ and $L$, the realization of the couplings (\ref{Coup}) is achieved  for 
\bea
d_{k,k+1}=\left[
\ln(\alpha/C_0)-\ln(\sqrt{k(N-k)})\right  ]\beta^{-1}.
\label{Dist2}
\eea
By means of directional couplers, we measured for our setup the dependence of the coupling on the separation of the waveguides at $\lambda=633$ nm, obtaining thus $\alpha\simeq 19.5$ cm$^{-1}$ and $\beta\simeq 0.152\,\mu$m$^{-1}$  (see Fig. \ref{fig1}). Using Eq.~(\ref{Dist2}), we were able to estimate the separations required for the realization of the coupling distribution (\ref{Coup}) in a PL of $N=9$ waveguides of length $L=10$ cm. Subsequently, the PL was inscribed in a fused silica glass using a standard femtosecond laser writing technique~\cite{Szameit2010a,Gattass2008}. The waveguides had an elliptical cross section ($4 \times 16$\,$\mu\textrm{m}^2$) and the associated refractive index modification was estimated to $\sim 5 \times 10^{-4}$ so that at $\lambda=633$ nm  a single mode is excited. Light propagation in the PL was monitored using a fluorescence microscopy technique (FMT)~\cite{Szameit2010a}. 
Following the approach outlined on pg. 7 of \cite{Szameit2010a}, propagation losses in the sample were estimated to about 0.4dB/cm, 
and the obtained fluorescence images, as well as the intensities at the input and the output of the sample, were normalized accordingly. 

\begin{figure}[t]
\centerline{\includegraphics[scale=0.828]{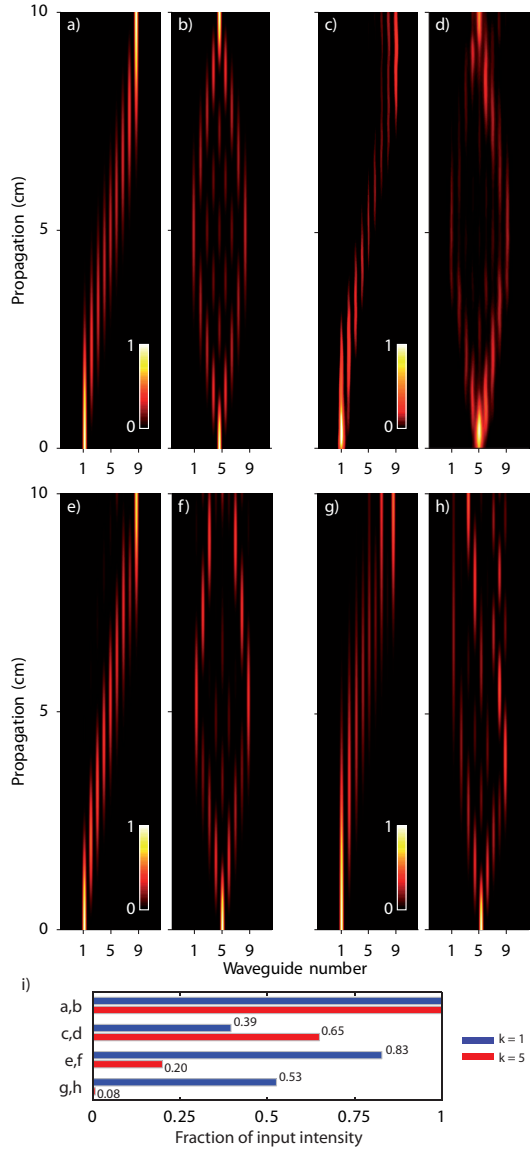}}
\caption{\label{fig2} (Color online) Light propagation in an array of $N=9$ waveguides with coupling distribution given by Eq. (\ref{Coup}). (a,b) Numerical results obtained by propagation of Eqs. (\ref{eq2}) when the 1st and the 5th waveguide are initially excited. (c,d) The corresponding fluorescence images obtained in our experiment. (e,f) As in (a,b) for $C_{k,k+1}\approx 0.56$~cm$^{-1}$. (g,h) As in (a,b) for $C_{k,k+1}=C_0\sqrt{k}$ and $C_0\approx 0.3$~cm$^{-1}$. (i) Fraction of the input intensity transferred from the $k$th 
to the $(N-k+1)$th waveguide.}
\end{figure}

%{\em Results and discussion.---}
Figures~\ref{fig2}(a) and~\ref{fig2}(b) present numerical results on the light propagation in the structure when one of the two outermost waveguides and the central waveguide respectively, are initially excited. The corresponding experimental observations are depicted in Figs.~\ref{fig2}(c) and~\ref{fig2}(d), and there is a rather good qualitative agreement with the theoretical predictions. Nevertheless, although propagation losses have been removed by rescaling, yet the power transfer from the $k$th to the $(N-k+1)$th waveguide is not complete, as opposed to the theoretical predictions. As shown in Fig. \ref{fig2}(i),  in the case of Fig. \ref{fig2}(c) about $39\%$ of the input intensity is transferred to the $9$th waveguide, whereas in the case of Fig. \ref{fig2}(d), this percentage increases to about $65\%$. In fact, a small fraction of the input power seems to have been transferred to waveguides adjacent to the $(N-k+1)$th waveguide; an indication that the diffraction of the signal has been restricted but it has not been minimized. Such deviations from the theory can be attributed to various experimental imperfections. (i) To follow the light propagation in the PL, a FMT was employed that  requires sufficiently intense writing laser. The optimal energy for waveguides of good  quality and intense enough fluorescence signal was 270 nJ, which is quite close to the threshold intensity for damaging the sample ($\sim$ 300 nJ). So, in view of the low repetition rate of the writing laser ($\sim$ 1kHz), small variations in the intensity during the writing process may have damaged locally some waveguides 
introducing unpredictable variations in the transverse and longitudinal modifications of the refractive index. This was the major source of disorder in our setup, leading to non-identical waveguides and a lattice without mirror symmetry. The use of a more appropriate laser source with higher repetition rate is expected to reduce considerably the writing time, facilitating thus the stability of the laser system and the induction of identical refractive index modifications. (ii) The theoretical predictions for perfect transfer rely on a well-defined coupling configuration, which can be achieved in our setup only within a certain accuracy. This is due to the translation stage in the writing procedure, which does not allow us to implement the desired waveguide separations with accuracy better than 0.5 $\mu$m. Our simulations show that such spacing inaccuracies have a small contribution to the observed deviations from theory, since they reduce the power transfer by at most 5\%; an estimate that can be reduced further by increasing the precision of the writing. The imperfection mechanisms and the propagation losses are expected to be present in the realization of  any coupling distribution in PLs (the details may vary though). 

Propagation losses are expected to be the same in our setup irrespective of the implemented coupling configuration and can be minimized~\cite{Fukuda2004}. By contrast, diffraction losses do depend on the realized coupling configuration. From Fig. \ref{fig2}(i), the diffraction losses in the implemented configuration were about 61\% and 35\% for Figs. \ref{fig2}(c) and (d), respectively. These losses would be significantly higher for configurations that are not judiciously designed, such as the uniform $C_{k,k+1}=C_0$ or the harmonic one $C_{k,k+1}= C_0\sqrt{k}$~\cite{Keil2011c}. As depicted in Fig. \ref{fig2}, in these configurations the transfer from the 1st to the 9th waveguide does not exceed ideally 83\%. This percentage drops with increasing $N$, as opposed to the ideally perfect transfer for (\ref{Coup}). Moreover self-imaging  is impossible when the central waveguide is initially excited; a distinct feature of configuration (\ref{Coup}) that has been verified experimentally [see Figs. \ref{fig2}(b,d)].

%{\em Limitations on the size of the lattice.---}
The NN approximation in our setup is justified, because the separations between successive waveguides are sufficiently large. According to Eq. (\ref{Dist2}), the minimum separations are for the central waveguides, which for $N\gg 1$ are practically equally spaced. From the exponential law previously discussed, the NN approximation is justified if  $e^{-\beta d_{\min}}=\epsilon$ for some finite $\epsilon\ll 1$, where $d_{\min}=d_{4,5}$. The parameter $\epsilon$ quantifies the deviations from the NN Hamiltonian; the larger $\epsilon$ is, the larger deviations we expect. For the implemented PL,  $d_{\min}=(21.9\pm0.5 )\mu$m and thus $\epsilon\lesssim 0.038$. For a given sample (i.e., fixed $L$ and $\alpha$), we can obtain a theoretical upper bound on $N$, so that the configuration (\ref{Dist2}) is implementable and couplings beyond NNs do not exceed a chosen $\epsilon$. With $d_{\min}$ given by Eq. (\ref{Dist2}) for $k=N/2$ we find $ N \leq 4 L \alpha \epsilon / \pi$, which implies that for a fixed $\alpha$, the design of larger networks requires longer samples. Propagation losses, however, increase exponentially with $L$ and thus, for large-scale all-optical networking one has to find the figure of merit between scalability and losses. In general, the presence of couplings beyond NNs does not preclude the existence of FC Hamiltonians~\cite{Kostak2007,Nik12}. 

%{\em Conclusions.---}
In conclusion, we have presented the photonic realization of a FC Hamiltonian with engineered couplings.  The implemented scheme is capable of minimizing losses associated with the diffraction of signals in linear PLs, allowing thus for reliable optical routing and switching when combined with the ideas of~\cite{Keil2011a,Nikolopoulos2008,Nik12}. Great effort is needed for minimization of disorder and imperfections, which are expected to affect the signal transmission. Our results provide a benchmark case and guide for the planning of future experiments on all-optical networking and short-distance quantum communication.

We acknowledge support by the EU Marie Curie Excellence Grant “MULTIRAD” Grant No. MEXTCT-2006-042683, and assistance of D. Gray in the experiments.

%%%%

\pagebreak
%\section*{Informational Fourth Page}
%In this section, we provide full versions of citations to assist reviewers and editors.

\end{document}